\documentclass[5p,12pt,sort&compress,preprint]{elsarticle}
\usepackage{makeidx}
\usepackage{amssymb}
\usepackage{psfrag}
\usepackage{soul}
\usepackage{amssymb,amsmath,amsthm}
\usepackage[utf8]{inputenc}
\usepackage{color}
\usepackage{graphicx}
\usepackage{amsmath,amsfonts,amsthm,amssymb,mathrsfs,bbm}
\usepackage[colorlinks=true,linkcolor=blue,citecolor=blue]{hyperref}
\usepackage[T1]{fontenc}
\usepackage{times}
\usepackage{multicol}

\journal{Chaos, Solitons \& Fractals}
\begin{document}

\begin{frontmatter}



\title{Dissipative structures  in a parametrically driven dissipative 
lattice: chimera, localized disorder, continuous-wave, and staggered state }

\author[UTA1]{A. M. Cabanas\corref{mycorrespondingauthor1}}
\cortext[mycorrespondingauthor1]{Corresponding author}
\ead{ana.cabanas.plana@gmail.com}
\author[UTA2]{J. A. Velez}
\author[UTA2]{L. M. P\'{e}rez}
\author[UFRO]{P. D\'{i}az}
\author[UCHILE]{M. G. Clerc}
\author[UTA2]{D. Laroze\corref{mycorrespondingauthor2}}
\cortext[mycorrespondingauthor2]{Corresponding author}
\ead{dlarozen@uta.cl}
\author[TelAviv]{B. A. Malomed}
\address[UTA1]{Sede Esmeralda, Universidad de Tarapacá, Av. Luis Emilio Recabarren 2477, Iquique, Chile}
\address[UTA2]{Instituto de Alta Investigaci\'{o}n, CEDENNA, Universidad de Tarapac\'{a}, Casilla 7D, Arica, Chile}
\address[UFRO]{Departamento de Ciencias F\'{i}sicas, Universidad de La Frontera, Casilla 54-D, Temuco, Chile}
\address[UCHILE]{Departamento de F\'isica  and Millennium Institute for Research in Optics, Facultad de Ciencias F\'isicas y Matem\'aticas, Universidad de Chile, Casilla 4873, Santiago, Chile}
\address[TelAviv]{Department of Physical Electronics, School of Electrical Engineering, Faculty of Engineering, and Center for Light-Matter Interaction, Tel Aviv University, Tel Aviv IL-69978, Israel}

\begin{abstract}
Discrete dissipative coupled systems exhibit complex behavior such as chaos, spatiotemporal intermittence, chimera among others.
We construct and investigate chimera states, in the form of confined
stationary and dynamical states in a chain of parametrically driven
sites with onsite damping and cubic nonlinearity.  The system is modeled by
the respective discrete parametrically driven damped nonlinear Schr\"{o}dinger equation. 
Chimeras feature quasi-periodic or chaotic dynamic in
the filled area, quantified by time dependence of the total norm (along with
its power spectrum), and by the largest Lyapunov exponent. Systematic
numerical simulations, in combination with some analytical results, reveal
regions in the parameter space populated by stable localized states of different
types. A phase transition from the stationary disorder states to spatially confined
dynamical chaotic one is identified. Essential parameters of the system are the strength and detuning of the forcing, 
as well as the lattice's coupling constant.
\end{abstract}

\end{frontmatter}

\section{Introduction}

\label{SS1}

Coupled oscillators are of great interest owing to their wide applicability in physics, chemistry, and biology 
(see Refs.~\cite{Kuramoto1984,Kivshar2004,Remoissenet2013, Ablowitz204, Kosevich1999, Kaneko1996}  and references therein).
Likewise, these systems attract global attention due to their dynamic behavior, such as
synchronization, defects and/or phase turbulence,
defect-mediated turbulence, spatiotemporal intermittency, and
coexisting coherent and incoherent states, among others \cite{Kuramoto1984, Kaneko1996,PikovskyKurths2002}.
In the last few decades, a great deal of effort has been devoted to understand
the coexistence of coherent and incoherent domains called chimera states.
These states were introduced by Kuramoto and Battogtokh \cite{Kura02}. 
This finding came as a
surprise because the oscillators were identical and symmetrically coupled.
Nevertheless, along with regular synchronized dynamic, the coupled
oscillator lattices exhibit incoherent or desynchronized behavior. The
understanding of the symmetry breaking, which leads to the emergence of
chimeras in dissipative systems \cite{Abrams04}, is a significant issue as
it appears in different contexts, such as biological models, delayed
systems, metamaterials, coupled map lattices, quantum systems, networks, and
even in a population of social agents, to mention a few \cite%
{Faghania18,Rybalova19,Clerc17,Kemeth16,Oleh2008,Laing2009,Oleh2010,Omelchenko2011,Wolfrum2011,Lee2011,Tinsley2012a,Tinsley2012b,Larger2013,Nkomo2013,Gautam2014,Panaggio2015,Berec2016,Santos17,Guo2018,Paras21,Haug21}.

As concerns dynamical elements of which lattices may be built, a
paradigmatic example of forced nonlinear oscillators is provided by the
parametrically driven damped nonlinear Schr\"{o}dinger equation \cite%
{Barashenkov91}. It models resonant phenomena in nonlinear dispersive media
and gives rise to soliton solutions with a variety of dynamical behavior
\cite{Alexeeva00,Barashenkov02,Barashenkov99,Zemlyanaya09}, Faraday waves
\cite{Coullet94}, breathers \cite{Barashenkov11a,Urzagasti13}, two-soliton
states \cite{Barashenkov11b,Urzagasti12}, and other soliton complexes \cite%
{Urzagasti14b}, and spatiotemporal chaos \cite{Shchesnovich02}.
Generalizations of this equation and its applications can be found in Refs.~\cite{Bara2003,CCL08,Burke2008,Malomed2009,CCL09a,CCL09b,CCL10a,CCL10b,Ma2010,CCL12,Urzagasti14a,Clerc15,Urzagasti14b,Leon17,Cabanas19,Urra19}.

In many physical realizations, this equation is replaced by its lattice
counterpart, in the framework of the class of models known as discrete
nonlinear Schr\"{o}dinger (DNLS) equations \cite{Kevrekidis09,new}. These
equations furnish fundamental models in discrete nonlinear optics \cite
{Lederer09,Kartashov09,Fleischer03,Yang05,Wang06,Kevrekidis05,Sakaguchi05,Chong09,Syafwan12}%
, as well as for Bose-Einstein condensates trapped and fragmented in deep
optical-lattice potentials \cite{Carre08,Kaurov05,Kaurov06,Smerzi03}. In
particular, the stability of discrete solitons in the parametrically driven
DNLS equations, both conservative and dissipative ones, has been studied in
Refs. \cite{Susanto06,Syafwan10,Syafwan13,Alfimov19a,Alfimov19b}.

The present work aims to study and  build up  chimera states in arrays of dissipative
parametrically driven coupled oscillators which are described by the damped
DNLS equation with the parametric gain applied at its sites. This state 
appear as patched-shaped (self-confined) patterns filled by
waves with a \textquotedblleft leaping" spatial structure (see an exact
definition below), which are linked by steep transient layers to the stable
zero state.  When the parameters are modified the chimera states 
are replaced by a spatially disordered localized one.
Such states exist in a stationary form due to the bistability and discrete nature of the system. 
In particular, we explore the existence of the chimera state as
a function of the driving force and its detuning, producing a complete chart
for stable stationary and dynamical states in the respective parameter
space. All these dynamic behavior correspond to permanent states of the 
system out of equilibrium. Which correspond to {\it dissipative structures} in the 
terminology introduced by Prigogine \cite{Prigogine1977}.
Effects of variation of the intersite-coupling strength in the
underlying DNLS equation are examined too. We find that, in addition to
being regular dynamical states, chimera, and the system gives rise to multistability.

The manuscript is organized as follows. The model and some analytical
findings for it are presented in Sec. \ref{SS2}. Results from systematic
numerical investigation of the stationary and dynamical states are summarized
in Sec. \ref{SS3}. Conclusions are presented in Sec. \ref{SS4}.

\section{The model and analytical results}

\label{SS2}

\subsection{The parametrically driven damped DNLS equation}

Let us consider a one-dimensional array of coupled nonlinear oscillators under
the action of dissipation and parametric drive. The respective DNLS equation
is (cf. Ref. \cite{Syafwan13})

\begin{gather}
\frac{dA^{j}}{dt}=-i\epsilon (A^{j+1}+A^{j-1}-2A^{j})  \notag \\
+\gamma \left( A^{j}\right) ^{\ast }-i\nu A^{j}-\alpha
A^{j}-i|A^{j}|^{2}A^{j},  
\label{A}
\end{gather}%
where $A^{j}$ is a complex-valued amplitude at site $j$, the asterisk symbol stands
for the complex conjugation, and $t$ is normalized time. If Eq. (\ref{A})
models an array of parametrically amplified lossy optical waveguides with
the Kerr nonlinearity \cite{array}, $t$, is actually, the propagation
distance along each waveguide. Further, $\epsilon >0$ represents the
coupling between adjacent sites in the array, $\gamma >0$ is the strength of
the parametric drive (the sign of $\gamma <0$ may be inverted by redefining $%
\tilde{A}^{j}\equiv iA^{j}$), $\nu $ is the detuning of the drive, and $%
\alpha >0$ is the damping constant, while the on-site nonlinearity
coefficient is scaled to be $1$. Note that the sign of the nonlinear term in
Eq.~(\ref{A}) corresponds to the self-focusing onsite nonlinearity; if its
original sign is opposite, it may be inverted by means of the staggering
transformation \cite{Cai}, that is, $A^{j}\equiv (-1)^{j}\exp \left( -2i\epsilon
t\right) \left( A^{j}\right) ^{\ast }$. The remaining scaling invariance of
Eq. (\ref{A}) allows us to fix one of the parameters. We use this option to
choose the value of the damping coefficient, $\alpha =0.35$, which is
convenient for the presentation of numerical results. The analysis is
reported below for negative detuning, $\nu <0$, which implies a possibility
of the existence of the dissipative bright discrete solitons, supported by the self-focusing
nonlinearity.

\begin{figure}[t]
\begin{center}
\includegraphics[width=0.4\textwidth]{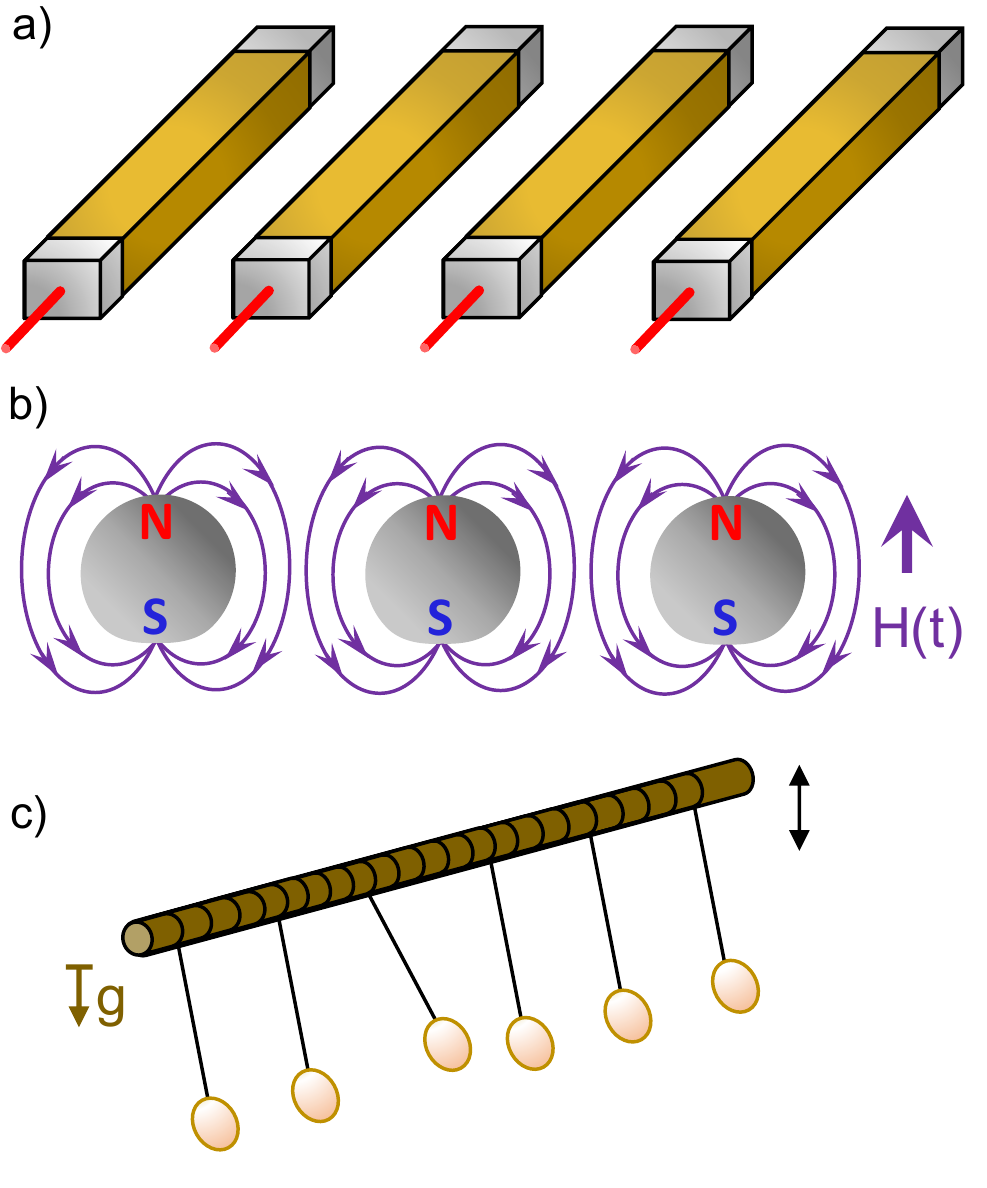}
\end{center}
\caption{Parametrically damped driven discrete systems. 
Schematic representation of  a one-dimensional  array of forcing coupled-waveguide resonators a),
coupled magnets under a time dependent magnetic field b), and vertically driven coupled pendulums c).}
\label{Fig-ParametricaSystems}
\end{figure}

It is relevant to mention that, in addition to the direct realizations of
Eq. (\ref{A}) as the chain of coupled oscillators with the complex
amplitudes, the same model can be derived as an asymptotic approximation for
a parametrically driven discrete nonlinear Klein-Gordon equation or Frenkel-Kontorava model, i.e., a
chain of nonlinear oscillators with real dynamical variables \cite{Alfimov19b}. 
Figure \ref{Fig-ParametricaSystems} illustrates some examples of different physical systems that are described by parametrically driven damped
discrete nonlinear Schr\"{o}dinger  equation (\ref{A}). Panel a) represents a one-dimensional  array of forcing coupled-waveguide resonators \cite{Clerc17}, frame b) illustrates a set of parametrically driven coupled magnets   \cite{CCL12,Urzagasti13,Urzagasti12}, while panel c) depicts a vertically driven coupled pendulums  \cite{CCL08}.

Stationary solutions to Eq. (\ref{A}) can be looked for with a constant
value of the phase at all sites:%
\begin{equation}
A^{j}=B^{j}\exp \left( \mp i\delta _{0}\right) ,\delta _{0}\equiv \tan
^{-1}\left( \sqrt{\frac{\gamma -\alpha }{\gamma +\alpha }}\right) .
\label{AB}
\end{equation}%
where real amplitudes $B^{j}$ obey the stationary version of the usual DNLS
equation,

\begin{gather}
\epsilon (B^{j+1}+B^{j-1}-2B^{j})+\left( B^{j}\right) ^{3}=  \notag \\
\left( -\nu \pm \sqrt{\gamma ^{2}-\alpha ^{2}}\right) B^{j}.  \label{B}
\end{gather}%
The top and bottom signs in Eq. (\ref{B}) correspond to those in Eq. (\ref%
{AB}).

In simulations reported in this work, we fixed the number of sites in the
lattice as $N=256$, so the discrete coordinate takes values $%
j=1,...,256 $. We used Neumann boundary conditions at edges of the lattice,
which means formally setting $A^{0}\equiv A^{1}$ and $A^{N+1}\equiv A^{N}$
in Eq. (\ref{A}) at $j=1$ and $j=256$, respectively.

\subsection{Stability of the zero state}

Because states considered in this work include zero-field segments, such
states are relevant solutions if $A^{j}=0$ is a stable solution of model Eq. (\ref%
{A}). To address the stability of zero, perturbed solutions to the
linearized version of Eq. (\ref{A}) are looked for as a combination of terms
$\exp \left( \Gamma t\right) \left\{ \cos \left( kj\right) ,\sin \left(
kj\right) \right\} $ to derive a dispersion relation for the instability
growth rate, $\Gamma $, as a function of the perturbation wavenumber $k$
\begin{equation}
\Gamma =-\alpha \pm \sqrt{\gamma ^{2}-\left[ 4\epsilon \sin ^{2}\left(
k/2\right) -\nu \right] ^{2}}  \label{Gamma}
\end{equation}%
(cf. Ref. \cite{Hennig}). It follows from Eq. (\ref{Gamma}) that stability
conditions for the zero state, $\mathrm{Re}\left( \Gamma (k)\right) \leq 0$,
which must hold for all real values of $k$, amount to%
\begin{eqnarray}
\gamma ^{2} &\leq &\alpha ^{2}+\nu ^{2}\equiv \gamma _{\max }^{2}, \quad~\mathrm{at}~\nu <0,  \notag \\
\gamma &\leq &\alpha , \quad  \quad \quad \quad \quad \quad \quad ~\mathrm{at}~0<\nu <4\epsilon ,  \notag  \label{gamma} \\
\gamma ^{2} &\leq &\alpha ^{2}+\left( \nu -4\epsilon \right) ^{2}, \quad~\mathrm{at}~\nu >4\epsilon . 
\end{eqnarray}%
In the continuum limit, which corresponds to $\epsilon \rightarrow \infty $,
or, effectively, to replacement
\begin{equation}
4\epsilon \sin ^{2}\left( k/2\right) \rightarrow \epsilon k^{2},
\label{limit}
\end{equation}%
in Eq. (\ref{Gamma}), the bottom line in Eq. (\ref{gamma}) is irrelevant,
while the top and middle ones remain valid, cf. Ref.~\cite{Coullet94}. In
the case of $\gamma ^{2}>\gamma _{\max }^{2}$~at $~\nu <0$ [see the top line
in Eq. (\ref{gamma})], the instability of the zero state leads to
establishment of solutions to Eq. (\ref{A}) in the form of Faraday patterns,
as \cite{Coullet94}.

\subsection{Continuous-wave states: existence conditions}

Next, Eq. (\ref{B}) gives rise to two uniform (\textquotedblleft
continuous-wave", CW)\ solutions with a constant amplitude:%
\begin{equation}
B=\sqrt{-\nu \pm \sqrt{\gamma ^{2}-\alpha ^{2}}}\equiv B_{0}^{\left( \pm
\right) }.  \label{CW}
\end{equation}%
Equation (\ref{CW}) also describes the amplitude of lattice solitons, with one
or several excited sites, in the \textit{anti-continuum limit}, $\epsilon
\rightarrow 0$ \cite{Aubry}.

In the case of positive detuning, $\nu >0$, solution (\ref{CW}) exists only
with the top sign, under condition $\gamma ^{2}>\alpha ^{2}+\nu ^{2}$, which
is incompatible with the middle and bottom lines in Eq. (\ref{gamma}), hence
in this case, the CW solution cannot coexist with the stable zero state. However, in the
case of $\nu <0$, CW solutions with both signs in Eq. (\ref{CW}) exist in
the interval of values of the drive's strength
\begin{equation}
\alpha ^{2}<\gamma ^{2}\leq \alpha ^{2}+\nu ^{2},  \label{<<}
\end{equation}%
in which the stability condition of the zero solution, given by the top line
in Eq. (\ref{gamma}), holds. Therefore, in this interval, both the zero state
and the CW solution with a larger amplitude, corresponding to the top sign
in Eq. (\ref{CW}), may be stable. Note that, according to general principles
of the bifurcation theory \cite{bif}, the latter solution is definitely
stable in the anti-continuum limit, with $\epsilon =0$ in Eq. (\ref{A}),
while the full stability analysis at $\epsilon >0$ makes necessary to
consider the possibility of the modulational instability, see the following
subsection. On the other hand, the intermediate solution, corresponding to
the bottom sign in Eq. (\ref{CW}), is definitely unstable, playing the role
of a \textit{separatrix} between attraction basins of the two presumably
stable states.

With the increase of $\gamma $, both CW states, given by Eq. (\ref{CW}),
emerge through the \textit{saddle-node bifurcation} \cite{bif} at $\gamma
=\alpha $. With the subsequent increase of $\gamma $, the unstable solution
with the smaller amplitude collides with the zero state, and thus
disappears, through the \textit{inverse pitchfork bifurcation }\cite{bif},
at $\gamma ^{2}=\alpha ^{2}+\nu ^{2}$ (formally, the latter bifurcation
involves solution $B_{0}^{(-)}$, given by Eq. (\ref{CW}), and its
counterpart $-B_{0}^{(-)}$). The same bifurcation destabilizes the zero
solution, as seen from the top line in Eq. (\ref{gamma}). At $\gamma
^{2}>\alpha ^{2}+\nu ^{2}$, only the CW solution with the top sign in Eq. (%
\ref{CW}) exists, remaining possibly stable, while the zero state is
unstable.

\begin{figure*}[!tbph]
\begin{center}
\includegraphics[width=0.85\textwidth]{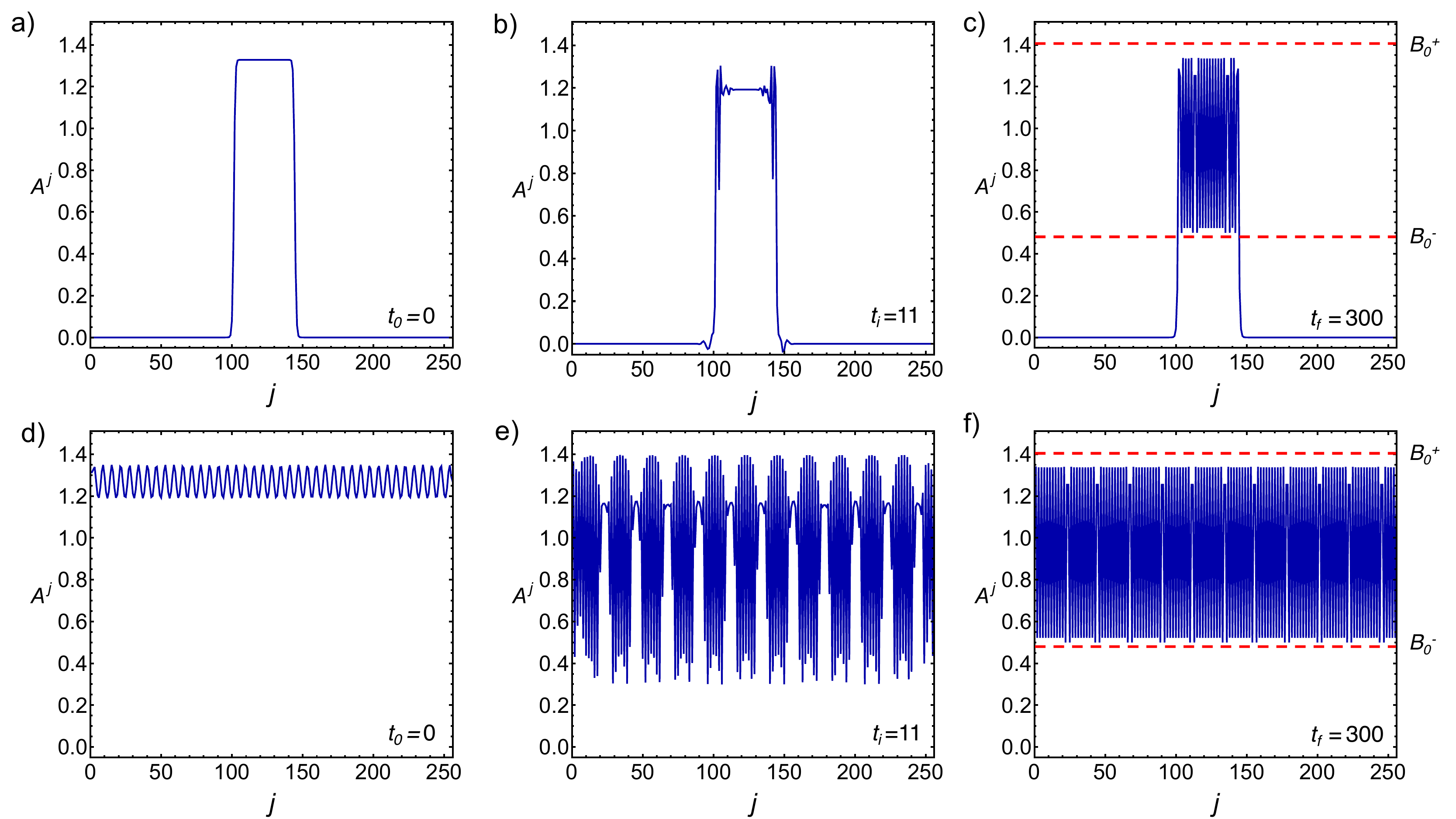}
\end{center}
\caption{Localized disorder sate. Panels a), b), and c) display the input, transient, and established
localized states, generated by simulations of Eq. (\protect\ref{A}) with $%
\protect\epsilon =0.51$, $\protect\alpha =0.35$, $\protect\gamma =1.01$, and
$\protect\nu =-1.15$ at three instants of time $t_{0}=0$, $t_{i}=11$, and $%
t_{f}=300$. Panels d), e), and f) display the corresponding solutions for
the extended state that occupies the entire spatial domain at the same value
of the parameters. Both the localized and extended states are filled by the
LW (\textit{leaping wave}). The red lines represent CW amplitudes described
by Eq (\protect\ref{CW}) for the given values of $\protect\alpha $, $\protect%
\gamma $, and $\protect\nu $. }
\label{figure1}
\end{figure*}

Thus, stable zero and CW states coexist in interval (\ref{<<}), under the
condition of $\nu <0$. Nevertheless, it is impossible to build up a
transient layer or domain wall (front solution) connecting these states, with exponentially
decaying tails approaching each one. Indeed, the tail approaching the zero
state can be looked for as the following solution to the linearized version
of Eq. (\ref{B})
\begin{equation}
B^{j}=b_{0}\exp \left( -q|j|\right),  \label{tail}
\end{equation}
at $|j|\rightarrow \infty $, with some constant $b_{0}$ and the spatial
decay rate, $q$, determined by expression%
\begin{equation}
\sinh ^{2}\left( q/2\right) =\left( 4\epsilon \right) ^{-1}\left( -\nu \pm
\sqrt{\gamma ^{2}-\alpha ^{2}}\right) ,  \label{sinh}
\end{equation}%
where the top and bottom signs correspond to those in Eqs. (\ref{CW}) and (%
\ref{tail}). On the other hand, the tail approaching the CW state is looked
for as
\begin{equation}
B^{j}=B_{0}^{\left( \pm \right) }+\tilde{b}_{0}\exp \left( -p|j|\right) ,
\label{layer}
\end{equation}%
with another constant $\tilde{b}_{0}$ and the decay rate $p$ determined by
Eq. (\ref{B}) linearized around $B=B_{0}^{\left( \pm \right) }$:%
\begin{equation}
\sinh ^{2}\left( p/2\right) =-\left( 2\epsilon \right) ^{-1}\left( -\nu \pm
\sqrt{\gamma ^{2}-\alpha ^{2}}\right) .  \label{p}
\end{equation}

Obviously, conditions $\sinh ^{2}\left( q/2\right) >0$ and $\sinh ^{2}\left(
p/2\right) >0$, which are necessary for the existence of the transient
layer, are incompatible, as it follows from Eqs. (\ref{sinh}) and (\ref{p}).
Therefore, localized patterns in the form of finite-width patches of CW
connected by smooth fronts to the stable zero state do not exist.
Nevertheless, we find that, at 
$\epsilon<0.44$, flat localized states filled by the CW do exist, being
bounded by layers which include sharp edge peaks, hence they cannot be
approximated by ansatz (\ref{layer}) \cite{Coullet2002, ClercFalcon2005}.

\begin{figure*}[!tbph]
\begin{center}
\includegraphics[width=0.75\textwidth]{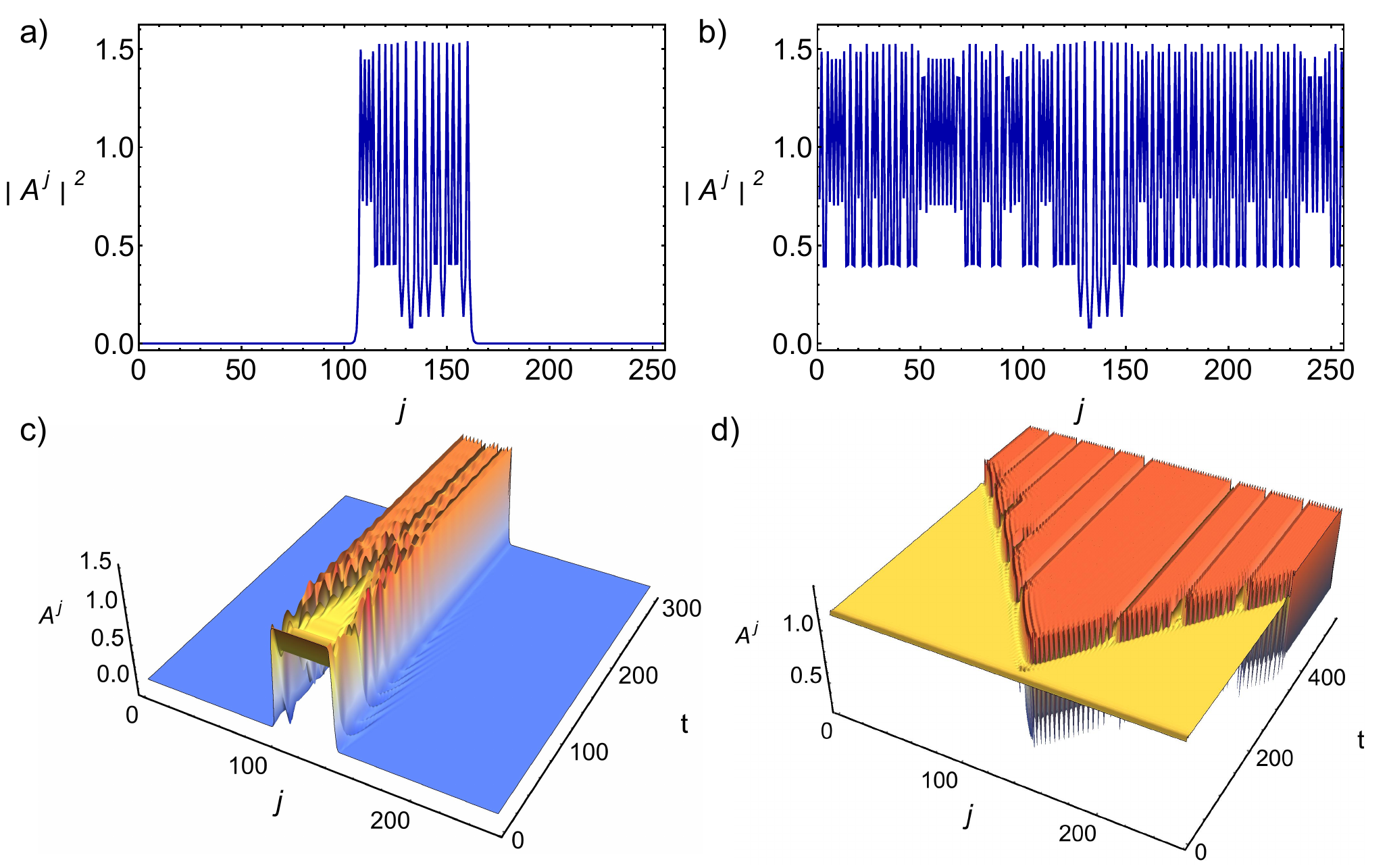}
\end{center}
\caption{Disorder localized state. Panels a) and b) show the squared absolute value, $|A^{j}|^{2}$,
for the {robust} confined localized state and the extended one,
respectively, filled by the LW (leaping wave). Panels c) and d) show the
spatiotemporal evolution of $\mathrm{Re}\left( A^{j}(t)\right) $ from the
flat-amplitude inputs towards the final states depicted in a) and b),
respectively. The solutions are produced by Eq. (\protect\ref{A}) at $%
\protect\gamma =0.55$, $\protect\nu =-1.0$ with $\protect\alpha =0.35$ and $%
\protect\epsilon =0.51$.}
\label{figure2}
\end{figure*}

Localized states produced by Eq. (\ref{A}) are the main
subject of this work. These states are built as localized patches filled not by flat CW
states, but rather by ones spatially oscillating between different amplitude
levels (\textit{leaping waves}, LWs). Chimera states are built as spatiotemporal incoherence segments
set on top of the stable zero background. Example of them are shown in
Figs. \ref{figure1} c), \ref{figure2} a) and \ref{fig2} below and 
explained in detail in section \ref{S4}.

In this framework, it is worth to mention that solutions for smooth
transient layers connecting zero and CW states, in spite of being impossible in the
present system, are admitted by models including competing nonlinearities.
Known examples are provided by equations including cubic-quintic \cite{Zeev}
or quadratic-cubic \cite{Marek} combinations of self-focusing and defocusing
terms.

\subsection{Modulational stability of the CW state}

It is also relevant to address the modulational (in)stability of the CW
state given by expressions (\ref{AB}) and (\ref{CW}). As it is well known, the
constant-amplitude solution of the usual DNLS equation with the
self-focusing sign of the onsite nonlinearity is always unstable, in the
absence of the damping and parametric drive \cite{Kivshar}. In the present
case, a perturbed form of the CW solution is sought for as
\begin{equation}
A^{j}(t)=B_{0}^{\left( \pm \right) }\left[ 1+a_{j}(t)\right] \exp \left[ \mp
i\delta _{0}+i\chi _{j}(t)\right] ,  \label{pert}
\end{equation}%
where $\mp $ has the same meaning as in Eq. (\ref{AB}). The substitution of
this expression in Eq. (\ref{A}) and the corresponding linearization with respect to small
perturbations, $a_{j}(t)$ and $\chi _{j}(t)$, leads to a system with the following evolution
equations for the perturbations,

\begin{align}
\frac{da_{j}}{dt }& =\epsilon \left( \chi _{j+1}+\chi _{j-1}-2\chi
_{j}\right)  \notag \\
& -2\gamma \sin \left( 2\delta \right)  \chi _{j},  \notag \\
\frac{d\chi _{j}}{dt }& =-\epsilon \left( a_{j+1}+a_{j-1}-2a_{j}\right)
\notag \\
& -2\left( B_{0}^{\left( \pm \right) }\right) ^{2}a_{j}-2\gamma \cos \left(
2\delta \right) \chi _{j}.  \label{14}
\end{align}

Eigenmodes of the small perturbations, with instability growth rate $\sigma $
(that may be complex-valued) and real wavenumber $k$, are looked for in the
usual form,%
\begin{equation}
\left\{ a_{j}(t),\chi _{j}(t)\right\} =\left\{ a^{(0)},\chi ^{(0)}\right\}
\exp (ikj+\sigma t).  \label{eigen}
\end{equation}%
The substitution of ansatz (\ref{eigen}) in Eq. (\ref{14}) yields a
dispersion equation for $\sigma (k)$
\begin{gather}
\sigma \left( \sigma +2\alpha \right) +\left[ 4\epsilon \sin ^{2}\left(
k/2\right) \mp 2\gamma \sin \left( 2\delta \right) +2\nu \right]  \notag \\
\times \left[ 4\epsilon \sin ^{2}\left( k/2\right) \mp 2\gamma \sin \left(
2\delta \right) \right] =0.  \label{sigma}
\end{gather}%
It is easy to see that the continuum limit of Eq. (\ref{sigma}), which is
defined as per Eq. (\ref{limit}), always gives rise to instability
(represented by a root with $\sigma >0$). On the other hand, the analysis
reveals a region in the parameter space in which Eq. (\ref{sigma}) secures
stability of the CW in the above-mentioned relevant case, with $\nu <0$ and
the top sign in Eqs. (\ref{AB}), (\ref{CW}), and (\ref{sigma})
\begin{equation}
\alpha ^{2}+4\epsilon ^{2}\equiv \left( \gamma _{\min }^{2}\right) _{\mathrm{%
CW}}<\gamma ^{2}<\alpha ^{2}+\nu ^{2}\equiv \gamma _{\max }^{2}.
\label{MI-stable}
\end{equation}%
The self-consistency condition for the double inequality in Eq. (\ref%
{MI-stable}) is $\epsilon <|\nu |/2$. Here the left inequality is the CW
modulational-stability condition proper, while the right one is copied from
the top line of Eq. (\ref{gamma}). Therefore, the CW state (\ref{AB}) with
amplitude $B_{0}^{(+)}$ and zero solution are \emph{simultaneously stable}
in the parameter \textit{window} defined by Eq. (\ref{MI-stable}). The
possibility of the respective \textit{bistability} is relevant to the
present work, which aims to build up  stable nonzero modes on top of the
stable zero background (although, as shown above, these two stable states
cannot be connected by a transient-layer solution).

On the other hand, straightforward consideration of Eq. (\ref{sigma})
demonstrates that, in the same case, the CW state with smaller amplitude,
which corresponds to the bottom sign in Eqs. (\ref{AB}), (\ref{CW}), and (%
\ref{sigma}), is always subject to the modulational instability. Another
obvious corollary of Eq. (\ref{MI-stable}) is that (as  it is well known \cite{Kivshar}) 
the stability interval (\ref{MI-stable}) does not exist in the
absence of the drive and damping, $\gamma =\alpha =0$.

\subsection{Staggered states}

\label{S4}

In addition to the stationary solutions with the fixed phase considered
above in the form of ansatz~(\ref{AB}), it is possible to introduce \textit{%
staggered solutions} of Eq. (\ref{A}), with alternating signs of amplitudes
at adjacent sites of the lattice, $A^{j}\equiv (-1)^{j}\tilde{A}^{j}$. This
transformation casts Eq. (\ref{A}) in the form of
\begin{gather}
\frac{d\tilde{A}^{j}}{dt}=i\epsilon (\tilde{A}^{j+1}+\tilde{A}^{j-1}+2\tilde{%
A}^{j})-i|\tilde{A}^{j}|^{2}\tilde{A}j  \notag \\
-i\nu \tilde{A}^{j}+\gamma \left( \tilde{A}^{j}\right) ^{\ast }-\alpha
\tilde{A}^{j}.  \label{staggered}
\end{gather}%
Further, the substitution of
\begin{equation}
\tilde{A}^{j}=\tilde{B}^{j}\exp \left( \mp i\delta _{0}\right) ,  \label{AC}
\end{equation}%
with the same $\delta _{0}$ as in Eq. (\ref{AB}) and real discrete field $%
\tilde{B}^{j}$, leads to the respective stationary equation,

\begin{gather}
-\epsilon (\tilde{B}^{j+1}+\tilde{B}^{j-1}+2\tilde{B}^{j})+\left( \tilde{B}%
^{j}\right) ^{3}=  \notag \\
\left( -\nu \pm \sqrt{\gamma ^{2}-\alpha ^{2}}\right) \tilde{B}^{j}.
\label{C}
\end{gather}%
Solutions to Eqs. (\ref{staggered}) and (\ref{C}) can be obtained from the
above unstaggered ones by substitution%
\begin{equation}
\epsilon \rightarrow -\epsilon ,\nu \rightarrow \nu -4\epsilon .  \label{eff}
\end{equation}%
In particular, a solution to Eq. (\ref{C}) with constant $\tilde{B}^{j}$
gives rise to the following staggered version of the CW state, which can be
obtained from the CW solution (\ref{CW}) via substitution (\ref{eff}) (the
solution is written here in terms of the original lattice field $A^{j}$,
rather than $\tilde{A}^{j}$):
\begin{equation}
A^{j}=(-1)^{j}\sqrt{4\epsilon -\nu \pm \sqrt{\gamma ^{2}-\alpha ^{2}}}\exp
\left( \mp i\delta _{0}\right) .  \label{stagg}
\end{equation}%
Two different solutions given by Eq. (\ref{stagg}) with opposite signs $\pm $
exist if, as above, condition $\gamma >\alpha $ holds, and the intersite
coupling strength is subject to constraint%
\begin{equation}
\epsilon >(1/4)\left( \nu +\sqrt{\gamma ^{2}-\alpha ^{2}}\right) .
\label{epsilon>}
\end{equation}%
In particular, for $\nu <0$ condition (\ref{epsilon>}) holds if $\gamma $
satisfies inequality (\ref{<<}), the same one which is necessary for the
coexistence of two unstaggered CWs given by Eqs. (\ref{AB}) and (\ref{CW}).

Further, staggered decaying-tail solutions of the linearized version of Eq. (%
\ref{C}), in the form of $\tilde{B}^{j}=\tilde{b}_{0}\exp \left(
-q|j|\right) $, which make possible to connect zero and nonzero states
[cf. Eq. (\ref{tail})], are obtained from Eq. (\ref{sinh}) with the help of
substitution (\ref{eff}):
\begin{equation}
\cosh ^{2}\left( q/2\right) =\left( 4\epsilon \right) ^{-1}\left( \nu \mp
\sqrt{\gamma ^{2}-\alpha ^{2}}\right) ,  \label{cosh}
\end{equation}%
while the stability conditions for the zero solution keep the form of Eq. (%
\ref{gamma}). In the case of $\nu <0$ (negative detuning), the tail solution
(\ref{cosh}) exists solely under condition $\gamma ^{2}>\alpha ^{2}+\nu ^{2}$%
, which makes the zero state unstable, as per the top line in Eq. (\ref%
{gamma}). Hence, the staggered CW solution cannot coexist with stable zero,
as parts of a chimera or localized state. In the case of positive detuning, $\nu >0$, a
possibility of building a chimera combining the stable zero and a stable
staggered CW is not promising either, because conditions necessary for the
existence of tail (\ref{cosh}), i.e., $\cosh ^{2}\left( q/2\right) >1$, and
of two CW states, as given by Eq. (\ref{stagg}), contradict each other.

Lastly, as a generalization of the staggered solutions, it is possible to
introduce \textquotedblleft twisted" ones (so named following Ref. \cite%
{Lederer}), in which stationary field $B^{j}$ [defined as in Eq. (\ref{AB})]
takes opposite values at sites separated by ones with $B^{j}=0$:%
\begin{equation}
B^{j}=\left\{
\begin{array}{c}
\sqrt{2\epsilon -\nu \pm \sqrt{\gamma ^{2}-\alpha ^{2}}},~j=4n, \\
0,~j=1+2n, \\
-\sqrt{2\epsilon -\nu \pm \sqrt{\gamma ^{2}-\alpha ^{2}}},~j=2\left(
1+2n\right) ,%
\end{array}%
\right.  \label{twisted}
\end{equation}%
cf. Eqs. (\ref{CW}) and (\ref{stagg}). In this work, we do not aim to
consider patterns of this type in detail.

\begin{figure*}[!tbph]
\includegraphics[width=\textwidth]{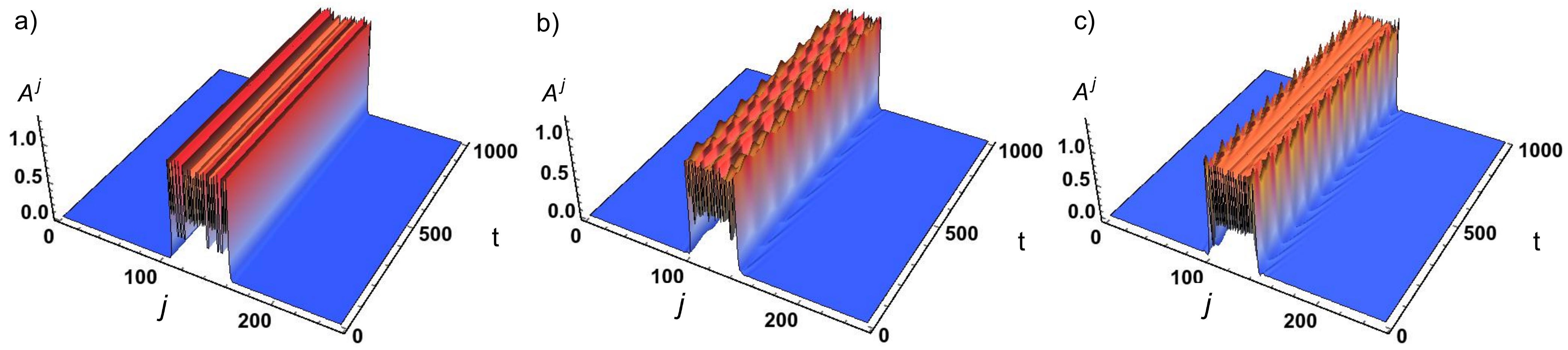}
\caption{Chimera states. Spatiotemporal evolution of $\mathrm{Re}\left( A^{j}(t)\right) $
for three species of self-confined chimera states, generated by simulations
of Eq. (\protect\ref{A}). (a) A disorder localized state, found for $\protect%
\gamma =0.67$, $\protect\nu =-1.0$, with $\protect\lambda _{\max }=-0.011$.
(b) A quasi-periodic chimera, for $\protect\gamma =0.75$ and $\protect\nu %
=-0.80$, with $\protect\lambda _{\max }=-5.54\times 10^{-5}$. (c) A chaotic
chimera, for $\protect\gamma =0.90$ and $\protect\nu =-1.2$, with $\protect%
\lambda _{\max }=0.07$.}
\label{fig2}
\end{figure*}

\section{Localized states}

\label{SS3}

\subsection{Localized disorder states}

The underlying equation (\ref{A}) with the above-mentioned boundary
conditions was solved by means of a variable-step fifth-order Runge-Kutta
scheme that ensures the relative precision of $10^{-7}$~\cite{Press92}. The
integration was performed up to $t=4.8\times 10^{3}$, which is essentially
larger than the sufficient time for the establishment of various static and
dynamical states. To identify the region of existence of chimera states, we
have first produced stable patterns filling the entire solution domain.
Then, the numerical solution produced chimeras by taking parameters from the
patterns' stability area, and running simulations of Eq. (\ref{A}) with
inputs in the form of a localized segment of the extended pattern, set in
the central part of the integration domain. The same localized states could be
produced from a simpler input, taken as a rectangular box with a constant
amplitude in a central segment of the integration domain, letting it to
evolve into a static or dynamical chimera of approximately the same width,
see an example below in Fig. \ref{fig2}(c).

The forcing, damping, and detuning terms play a crucial role in the
formation of various states in the present model. As said above, the damping
coefficient is fixed to be constant through rescaling, $\alpha =0.35$,
therefore results were collected by varying $\gamma $ and $\nu $, while (in
most cases) the intersite coupling constant was kept constant, at value $%
\epsilon =0.51$, which allows to present generic findings.

Figure \ref{figure1} shows examples of stationary states, produced as
numerical solutions of Eq. (\ref{A}) with $\gamma =1.01$ and $\nu =-1.15$.
Panel a) displays the above-mentioned input, taken as a rectangular box with
a flat amplitude. Panel b) represents a transient state produced in the
course of the evolution, that leads to the establishment of the stationary
localized disorder depicted in panel c). It is a finite-length patch of a quasi-regular
LW separated by steep edges (fronts) from the zero state. This localized disorder state is
stable  and motionless at the present values of the parameters, according to the top line in
Eq. (\ref{gamma}).

Further, panel f) displays a stationary LW pattern occupying the entire
integration domain, obtained via an intermediate state (panel e)) from the
input shown in panel d). It was introduced as a small-amplitude periodic
wave with the mean value close to the CW amplitude $B_{0}^{+}$, given by Eq.
(\ref{CW}) for current values of the parameters: $B_{0}^{\left( +\right)
}\approx 1.448$ and $B_{0}^{\left( -\right) }\approx 0.450$. The periodic
component was added to the input to initiate the development of the extended
LW state. Note that both values $B_{0}^{\left( \pm \right) }$ are clearly
visible as the limits, shown by red lines in panels c) and f), between which
the LW oscillates in both the confined (localized disorder state) and extended states. The
numerical data demonstrate that the stationary solutions displayed in panels
c) and f) have a spatially uniform phase.

Similar dynamic is observed at other values of the parameters. In
particular, Fig. \ref{figure2} displays the full picture of the
establishment of the patch-shaped (localized disorder state) state with the LW structure, and
its counterpart filling the entire domain, at $\gamma =0.55$, $\nu =-1.0$, $%
\alpha =0.35$, and $\epsilon =0.51$. Figures \ref{figure2} c) and d)
corroborate the modulational instability of the above-mentioned CW state,
with $B_{0}^{\left( +\right) }\approx 1.448$, in the confined and fully
extended forms alike, in agreement with the fact that the current parameter
values do not satisfy the left-side inequality in the CW stability condition
given by Eq. (\ref{MI-stable}). The observation that the instability onset
and development are essentially the same in panels c) and d) is explained by
the fact that the instability maximum determined by Eq. (\ref{sigma})
corresponds to $k=\pi $, hence the corresponding perturbation wavelength, $%
2\pi /k=2$, is much smaller than the width of the initial CW domains in both
panels.

Another essential conclusion suggested by Figs. \ref{figure1} and \ref%
{figure2} is that a sharp edge separating the zero and nonzero parts of the
finite-size patch (localized stated) is always stable, which makes possible to
build up a stationary localized disorder state of an arbitrary size, confined by a pair of
edges. Thus, the system demonstrates multistability, including localized states of
different sizes, as well as configurations with two or several localized domains
separated by zero-amplitude segments (not shown here).


\begin{figure}[h]
\begin{center}
\includegraphics[width=0.45\textwidth]{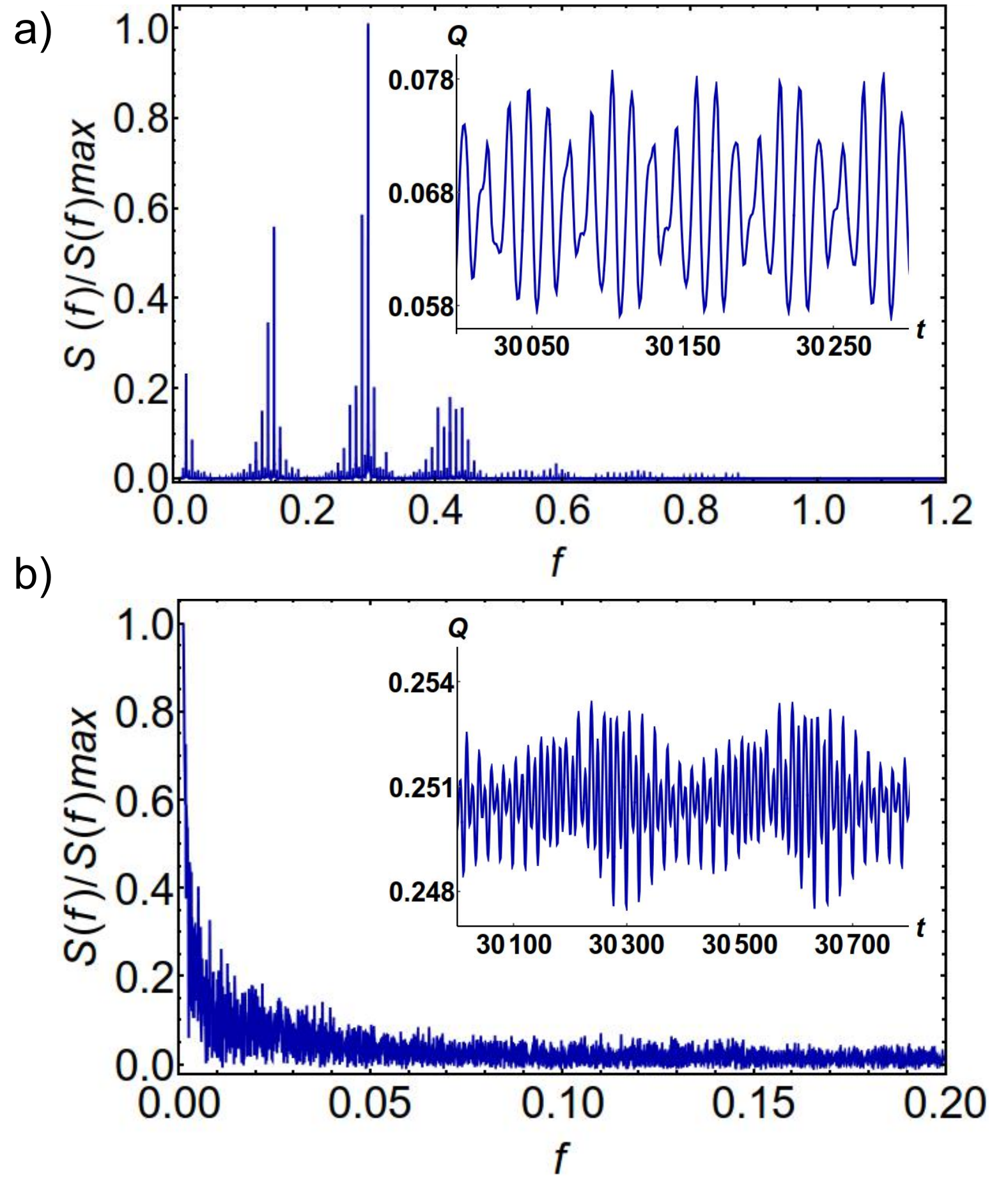}
\end{center}
\caption{The power spectrum, $S(f)$, of the time-dependent total norm, $Q(t)$%
, defined as per Eq. (\protect\ref{power}). Panels a) and b) correspond to
the quasi-periodic and chaotic chimera, displayed
in Figs. \protect\ref{fig2} b) and c), respectively. The corresponding time dependences $Q(t)$
[see Eq. (\protect\ref{Eq2})] are displayed in the insets.}
\label{fig3}
\end{figure}

\subsection{ Dynamical indicators}

To characterize different nature of dynamical behavior of localized states, we use, first, the scaled
total norm, which (along with the Hamiltonian) is a dynamical invariant of
the DNLS equation in the absence of the damping and drive \cite{Kevrekidis09}%
, while in the present case it is a function of time, in non-stationary
states
\begin{equation}
Q(t)=\frac{1}{N}\sum_{j=1}^{N}|A^{j}(t)|^{2}.  \label{Eq2}
\end{equation}%
In the case of time-dependent $Q(t)$, we computed its power spectrum,
\begin{equation}
S(f)={|\mathfrak{F}(f)|}^{2},  \label{power}
\end{equation}%
as a function of frequency $f$ [related to the angular frequency as $%
f=\omega /\left( 2\pi \right) $], where the Fourier transform of $Q(t)$ is
defined as
\begin{equation}
\mathfrak{F}(f)=\frac{1}{\sqrt{2\pi }}\int_{0}^{t_{\max }}Q(t)\exp \left(
-ift\right) dt.
\end{equation}%
A general principle is that ${S(f)}$ features a quasi-discrete spectrum,
with a set of narrow peaks, for regular solutions, which exhibit (quasi-)
periodic evolution in time. On the other hand, the spectrum is expected to
be an essentially continuous one if the underlying time-dependent solution
can be of chaotic  nature \cite{Ott}, see also a recent realization of the principle for
the nonlinear Schr\"{o}dinger equation with a trapping potential \cite{Tom}.

Another relevant indicator characterizing the dynamical behavior is the
\textit{largest Lyapunov exponent} (LLE), $\lambda _{\max }$ \cite%
{Ott,Wolf85}. When the system is chaotic, two trajectories, that were
infinitesimally close initially, separate in the phase space exponentially,
the distance between them growing $\sim \exp \left( \lambda _{\max }t\right)
$, with $\lambda _{\max }>0$. On the other hand, close trajectories converge
to the same non-chaotic attractor if $\lambda _{\max }<0$. The marginal case
of $\lambda _{\max }=0$ implies that the behavior is time-periodic,
quasi-periodic, or complex behavior with power sensitivity to initial conditions. 
LLE is widely used to quantify regular and chaotic evolution
in diverse dynamical systems, see, e.g., Refs. \cite%
{Wolf85,Eckmann86,Cross06,Geist90,Gallas10,Karimi12,Sprott2003,Mag12,Mag13}.


\begin{figure*}[!tbph]
\par
\begin{center}
\includegraphics[width=0.75\textwidth]{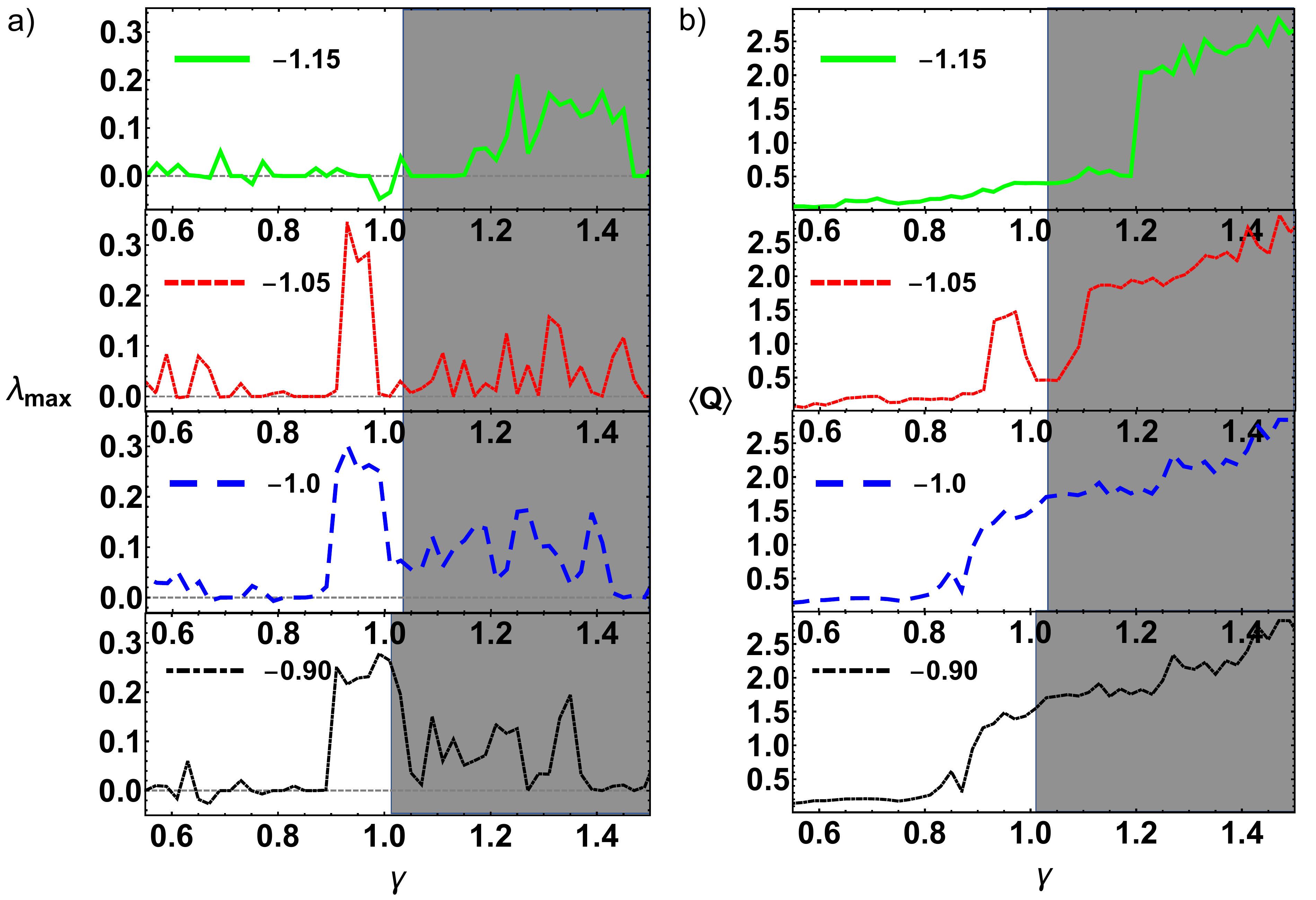}
\end{center}
\par
\caption{Dynamical indicators. a) LLE and b) time-average of the total norm $\left\langle
Q\right\rangle $ as functions of $\protect\gamma $ for fixed negative values
of $\protect\nu $, which are indicated in the panels. Other parameters in
Eq. (\protect\ref{A}) are $\protect\mu =0.35$, $\protect\epsilon =0.51$, and
$\protect\alpha =0.35$. In shaded areas the stationary LW (leaping wave),
filling the entire lattice, is unstable. }
\label{fig4}
\end{figure*}

\subsection{Chimera  states}

To collect systematic numerical data, we ran simulations of model Eq. (\ref{A})
with the initial state in the form of a state with a flat amplitude, filling
a central segment of the integration domain, as shown above in Figs. \ref%
{figure1} a) and \ref{figure2} c). Generic results, obtained at different
values of the driving strength, $\gamma $, and negative detuning, $\nu $,
are displayed in Fig. \ref{fig2}, in the form of established dynamical
regimes, observed in the course of a long simulation interval, $0<t<1000$.
All these self-sustained states were created in a finite lattice segment, $%
100<j<160$.

In panel \ref{fig2} a), the evolution at $\gamma =0.67$ and $\nu =-1.0$
leads to the establishment of a stable time-independent localized disorder state. In
this case, the respective value given by the top line in Eq. (\ref{gamma})
is $\gamma _{\max }\approx \allowbreak 1.06$, hence the zero background is
indeed stable. The regular {behavior} of this localized state  as a whole is
corroborated by the computation of the respective LLE, which is found to be {%
negative}, $\lambda _{\max }=-0.011$.

Next, increasing the strength of the forcing, a chimera featuring quasi-periodic temporal dynamic is displayed in
panel \ref{fig2} b) for $\gamma =0.75$ and $\nu =-0.80$, which is closer to
the respective stability boundary for the zero state, \textit{viz}., $\gamma
_{\mathrm{\max }}\approx 0.87$, as per the top line in Eq. (\ref{gamma}). 
Notice that this solution accounts for a coexistence of coherent and incoherent domains.
In this case, LLE is $\lambda _{\max }=-5.54\cdot 10^{-5} \pm 5.72 \cdot 10^{-6}$, which is actually a
numerical zero, as may be expected for quasi-periodic states. Finally, a
chaotic chimera, featuring apparently random spatiotemporal dynamic at $%
\gamma =0.90$ and $\nu =-1.2$ in a self-confined segment of the lattice, is
displayed in panel c). In the latter case, Eq. (\ref{gamma}) yields $\gamma
_{\max }\approx 1.25$, hence the zero background remains stable. The
numerically computed LLE is positive, $\lambda _{\max }=0.07$, which clearly
corroborates the chaotic character of the state as a whole \cite{Ott}.

\begin{figure}[h]
\includegraphics[width=0.45\textwidth]{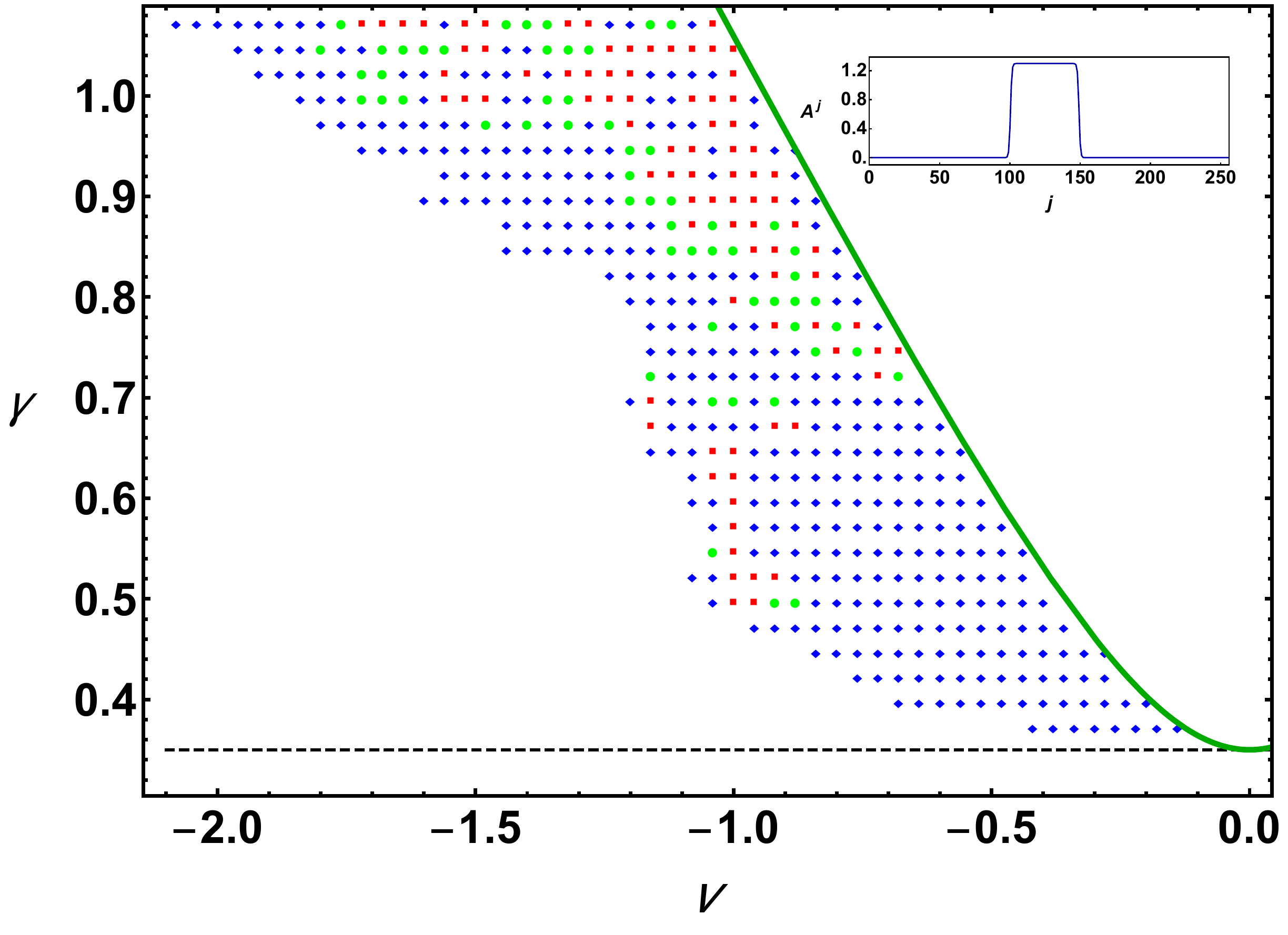}
\caption{Different species of robust self-confined states, generated by
simulations of Eq. (\protect\ref{A}) with the same localized input (shown in
the inset), are chartered in the parameter plane of the detuning, $\protect \nu $, and parametric-drive's strength, $\protect\gamma $, for a fixed
inter-site coupling constant, $\protect\epsilon =0.51$, and the value of the
dissipation coefficient fixed by the rescaling, $\protect\alpha =0.35$. Blue
diamonds, green circles, and red squares represent, respectively, stable
localized disorder state, chimera with quasi-periodic evolution, and
chaotic spatiotemporal dynamic in the spatially confined area.
The dashed horizontal line, $\protect\gamma =\protect\alpha \equiv 0.35$, is
the boundary above which two CW states (\protect\ref{CW}) exist at $\protect%
\nu <0$. Only the CW state may exist, as a stable one, above the green
continuous curve, i.e., at $\protect\gamma >\protect\gamma _{\max }=\protect%
\alpha ^{2}+\protect\nu ^{2}$ [see the top line in Eq. (\protect\ref{gamma}%
)], where the zero solution and, hence, any finite-width chimera as a whole
are unstable. Indeed, for $\protect\alpha =0.35$ and $\protect\epsilon -0.51$%
, the stable CW exists at $\protect\gamma >\protect\sqrt{\protect\alpha %
^{2}+4\protect\epsilon ^{2}}\approx 1.08$ [as per the left-hand inequality
in Eq. (\protect\ref{MI-stable})], i.e., above the area displayed in the
present figure. No stable nonzero states are produced by the simulations in
the white area to the left of the filled region.}
\label{fig5}
\end{figure}

To gain deeper understanding of the dynamic, the quasi-periodic and chaotic
character of the chimeras displayed in Figs. \ref{fig2} b) and c) is
quantified by the corresponding power spectra, defined as per Eq. (\ref%
{power}), which are presented in Figs. \ref{fig3} a) and b), respectively.
As expected, the spectrum is quasi-discrete for the quasi-periodic
oscillations and continuous for the chaotic dynamic.
The previous analysis allows us to conclude that the emergence of quasi-periodic and chaotic 
chimeras follows a route of extended quasi-periocity \cite{ClercVerschueren2013}.

The effect of the strength of the parametric forcing, $\gamma $, on
properties of the states under the consideration is represented by Fig. \ref%
{fig4}, which shows LLE and the time-average value of the total norm,
\begin{equation}
\left\langle Q\right\rangle =t_{\max }^{-1}\int_{0}^{t_{\max }}Q(t)dt,
\label{Q}
\end{equation}%
as functions $\gamma $, at different fixed values of detuning $\nu $, for
the self-confined states created in the above-mentioned lattice segment, $%
100<j<160$. The transition from regular to chaotic dynamic with the
increase of the forcing strength, $\gamma $, i.e., from $\lambda _{\max
}\approx 0$ to definitely positive values of $\lambda _{\max }$, coincides
with the transition from nearly constant to growing values of $\left\langle
Q\right\rangle $. Both phenomena are manifestations of the phase transition
from regular dynamic to chaos, which takes place at a critical point, $%
\gamma =\gamma _{\mathrm{crit}}$. For example, $\gamma _{\mathrm{crit}}(\nu
=-0.90)\approx 0.92$ and $\gamma _{\mathrm{crit}}\left( \nu =-1.05\right)
\approx 0.90$, while the respective values given by Eq. (\ref{gamma}) are $%
\gamma _{\max }\left( \nu =-0.9\right) \approx 0.966$ and $\gamma _{\max
}\left( \nu =-1.05\right) \approx 1.107$ [recall we have fixed the
dissipation constant in Eq. (\ref{A}) as $\alpha =0.35$, by means of
scaling]. Thus, the transition to the chaotic dynamic happens when the zero
state is still stable, at $\gamma <\gamma _{\max }$. Generally, ratio $%
\gamma _{\mathrm{crit}}/\gamma _{\max }$ takes values between $0.81$ and $%
0.90$.

Moreover, in the shaded area in Fig. \ref{fig4} direct
simulations demonstrate that the extended LW state is unstable. Thus, in an
interval of the drive's strength between $\gamma _{\mathrm{crit}}$ and the
left edge of the shaded area the dynamic is categorized as chaotic, while
the explicit LW instability does not set in, as yet. The \textquotedblleft
delay" of the onset of the instability after the transition to the positive
LLE may be a consequence of the finite width of the self-confined patch
(localized state) filled by LW. Finally, stable CWs are easily found at $\gamma
>\left( \gamma _{\min }\right) _{\mathrm{CW}}$, where, for the current
values of the parameters, $\alpha =0.35$ and $\epsilon =0.51$, the left-hand
inequality in Eq. (\ref{MI-stable}) gives $\left( \gamma _{\min }\right) _{%
\mathrm{CW}}\approx 1.08$.

\begin{figure}[tbp]
\begin{center}
\includegraphics[width=0.45\textwidth]{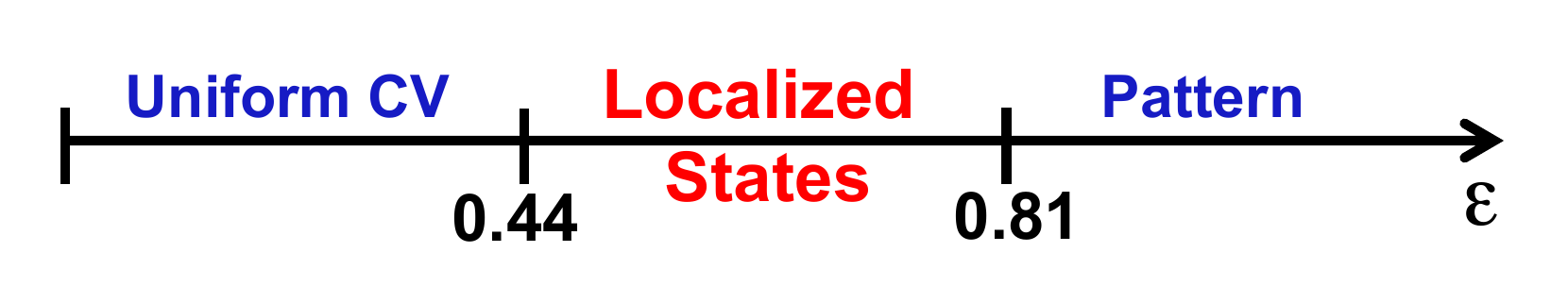}
\end{center}
\caption{Existence regions of different stable states as function of the
intersite coupling constant, $\protect\epsilon $, at fixed values of the
drive's strength, $\protect\gamma =0.95$, and detuning, $\nu =-1.11$.
The localized states (localized disorder and chimeras) are stable in the interval of $0.44<\protect%
\epsilon <0.81$. At $\protect\epsilon <\protect\epsilon _{\mathrm{crit}%
}\approx 0.44$ [see Eq. (\protect\ref{crit})], stable states, depending on 
the initial conditions, are uniform CWs or broad flat patches of CWs,
bounded by narrow
transient layers with sharp (but low) peaks, see Fig. \protect\ref{fig8}. 
At $\protect\epsilon >0.81$, Eq. (\protect\ref{A})
gives rise to stable delocalized patterns with zero crossings (not
considered here in detail).}
\label{fig6}
\end{figure}

\begin{figure*}[!tbph]
\begin{center}
\includegraphics[width=0.95\textwidth]{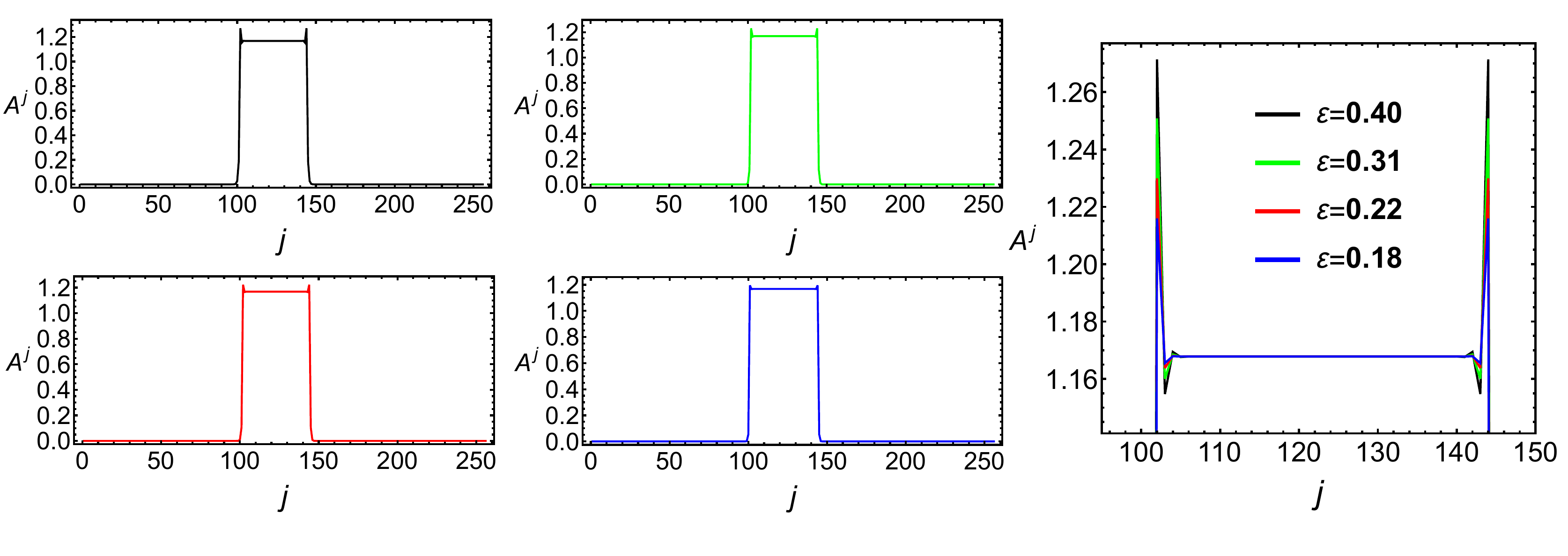}
\end{center}
\caption{$\mathrm{Re}\left( A^{j}\right)$ for stable states in the
form of confined flat (CW) patches, bounded by narrow layers with
sharp peaks. The states are obtained at $\gamma =0.95$, $\nu =-1.11$, and 
$\alpha =0.35$ (the same as in Fig. \ref{fig6}),
with $\epsilon =(0.40,0.31,0.22,0.18)$, as
color coded in the right panel (which provides a zoom of
the sharp peaks). Note that all these values of the lattice
coupling constant belong to the region (\protect\ref{crit}).}
\label{fig8}
\end{figure*}

Results of the systematic numerical analysis are summarized in Fig. \ref%
{fig5}, which is a chart designating different established states in the
parameter plane of $\left( \nu ,\gamma \right) $. The states are
categorized, with the help of the value of LLE, $\lambda _{\max }$, and the
character of the power spectrum of $Q(t)$ (quasi-discrete or continuous), as
stationary, temporarily quasi-periodic, or chaotic self-confined chimeras
(blue diamonds, green circles, and red squares, respectively, in Fig. \ref%
{fig5}). The results were generated by means of sufficiently long
simulations of Eq. (\ref{A}), initiated, at all values of $\nu $ and $\gamma
$, by the rectangular excitation field displayed in the inset to Fig. \ref%
{fig5}. The localized states, which include extended segments of the stable zero
solution, exist between the dashed horizontal line, $\gamma =\alpha \equiv
0.35$, and the solid green curve, $\gamma =\gamma _{\max }$ [see the top
line in Eq. (\ref{gamma})]. These boundaries correspond, respectively, to
the above-mentioned saddle-node and inverse pitchfork bifurcations. In
particular, at values of the drive's strength $\gamma <0.48$ and $%
0.55<\gamma <0.65$ for $\nu >-1.0$ the chart contains only stationary
chimeras.

We stress that the model obviously features multistability, as different
species of the stable state coexist at the same values of the parameters. In
particular, as{\ mentioned above (see Figs. \ref{figure1}, \ref{figure2}, %
\ref{fig2}), Eq.~(\ref{A}) admits bistability between 
localized and delocalized states, which are readily found as
stable solutions at the same values of the parameters, the choice between
them being determined by the initial conditions.}

\begin{figure*}[!tbph]
\begin{center}
\includegraphics[width=0.7\textwidth]{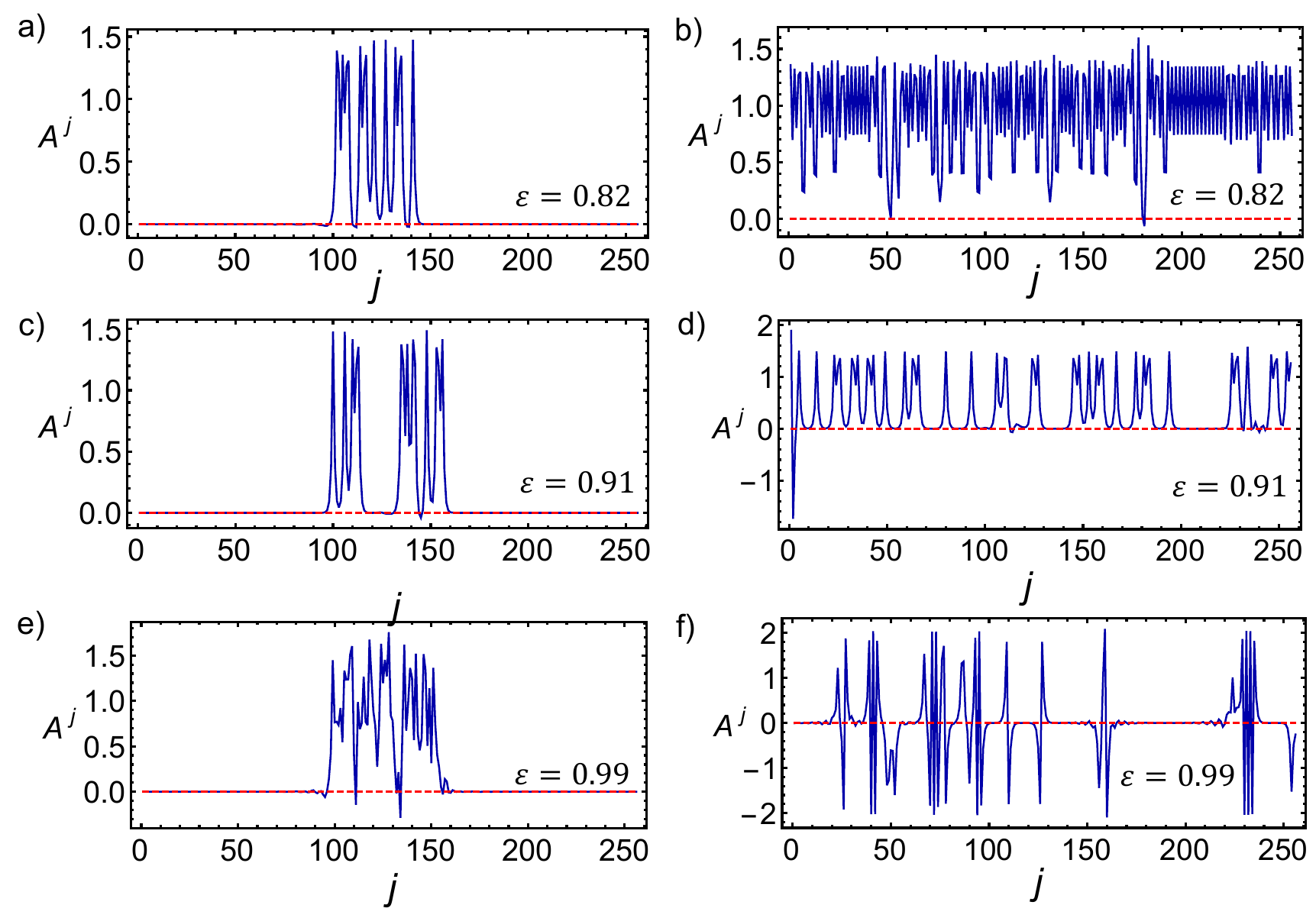}
\end{center}
\caption{Stable randomly shaped patterns produced by the simulations at the same 
fixed values of the parameters, $\gamma =0.95$, $\nu =-1.11$, and
$\alpha =0.35$, as in Figs. \ref{fig6} and \ref{fig8}. The patterns are 
produced at gradually increasing values of the lattice coupling constant, 
$\epsilon$, which leads to deeper zero crossing. The left and right columns
display, severally, confined and extended patterns, produced by respective 
inputs.}
\label{fig9}
\end{figure*}

The above results were obtained for the fixed value of the intersite
coupling constant, $\epsilon =0.51$. It is also relevant to consider
effects of variation of $\epsilon $. In Fig. \ref{fig6} we report the
results for fixed $\gamma =0.95$ and $\nu =-1.11$, the respective value
given by the top line of Eq. (\ref{gamma}) being $\gamma _{\max }\approx
\allowbreak 1.16$. The figure demonstrates that, at $\epsilon <0.44$, the
simulations produce no chimeras, which are ousted by the stable uniform CW, in
exact agreement with the left-hand inequality in Eq. (\ref{MI-stable}),
which shows that CWs are stable at
\begin{equation}
\epsilon <\epsilon _{\mathrm{crit}}\equiv (1/2)\sqrt{\gamma ^{2}-\alpha ^{2}}%
\approx 0.44,  \label{crit}
\end{equation}%
for $\gamma =0.95$ and $\alpha =0.35$. Detailed simulations demonstrated 
that, in the same region at 
$\epsilon <\epsilon _{\mathrm{crit}}$, the established numerical solution
depends on initial conditions. If an extended pattern is taken as the
input, simulations establish a uniform CW with amplitude $B_0^{+}$, see Eq. 
(\ref{CW}). On the other hand, the input in the form of a 
localized region filled by the LW gives rise to robust
solutions in the form of wide but confined flat CW patches, which are bounded
by narrow transient layers including sharp (but relatively low) 
peaks, such as the ones shown in Fig.
\ref{fig8}. Because of the presence of the peaks, these states cannot be predicted by ansatz (\ref{layer})
considered above. The right panel on Fig. \ref{fig8} demonstrates that the
height of the peaks diminishes with the decrease of $\epsilon $.

As shown in Fig. \ref{fig6}, for the same fixed values $\gamma =0.95$ and $%
\nu =-1.11$, stable finite-width chimeras are produced by numerical
simulations in an interval of values of the coupling constant $0.44<\epsilon
<0.81$. Lastly, at $\epsilon >0.81$, the former chimera patterns acquire zero
points (originally falling to $A_j=0$, and then crossing zero), as shown 
in Fig. \ref{fig9}. However, these states are different from the zero-crossing 
twisted ones, given by Eq. (\ref{twisted}), as they are apparently randomly built 
patterns. As shown in Fig. \ref{fig9}, they may feature both confined and 
extended shapes, depending on the input.

\section{Conclusion}

\label{SS4}

In this work, we have presented chimera
states and localized disorder state found in the parametrically driven discrete dissipative system,
modeled by the damped DNLS equation with the onsite parametric drive. The
system can be implemented as an array of optical fibers in which losses are
compensated by the parametric amplification. The localized states are built as
finite segments filled by a pattern which may be a stationary LW (a
spatially \textit{leaping wave} oscillating between lattice sites), or,
depending on parameters, an LW featuring quasi-periodic or chaotic time
dependence. The segment is connected by steep fronts to stable zero states
(in this sense, it is a chimera combining stable nonzero and zero modes).
The dynamic of the chimeras is characterized by the time dependence of the
total norm, and by LLE (the largest Lyapunov exponent). Our analysis reveals
a specific region of parameters in which robust localized disorder  and
 chimeras state exist, being the
strength and detuning of the parametric drive the essential control parameters. The dependence of the
existence region on the strength of the intersite coupling, $\epsilon $, is
considered as well. Besides, for smaller values of $\epsilon$,
stable states are represented by flat CW patches, which are confined by
narrow transient layers containing sharp peaks. A part of the numerical
findings is explained by analytical results that address the existence and
stability of the zero and uniform CW states.

As an extension of the analysis, it may be relevant to consider chimera
states in the two-dimensional version of the present system. Work in this
direction is in progress.

\section*{Acknowledgments}

LMP, PD, and DL acknowledge partial financial support from FONDECYT 1180905. DL
acknowledges the partial financial support from Centers of excellence with
BASAL/CONICYT financing, Grant AFB180001, CEDENNA. The work of BAM is
supported, in a part, by the Israel Science Foundation, through grant No.
1286/17. This author appreciates hospitality of Instituto de Alta Investigaci%
\'{o}n at Universidad de Tarapac\'{a} (Arica, Chile).
MGC thanks for the financial support of FONDECYT projects 1180903 and 
Millennium Institute for Research in Optics ANID-Millennium Science Initiative Program--ICN17\_012.

\end{document}